\ifx\destination\undefined
  \def\destination{arxiv}  
\fi
\def\arxiv{arxiv}
\def\aanda{aanda}
\def\publisher{publisher}
\RequirePackage{snapshot}
\PassOptionsToPackage{dvipsnames}{xcolor}
\PassOptionsToPackage{colorlinks=True,allcolors=RoyalBlue}{hyperlink}

\ifx\destination\arxiv
  \documentclass{article}
  \usepackage{aux/preprint}
  \usepackage{natbib}
  \usepackage{authblk}
  \let\address\affil
\fi
\ifx\destination\aanda
  \documentclass{aa}
  \usepackage{natbib}
  \bibpunct{(}{)}{;}{a}{}{,} 
\fi
\ifx
  \documentclass[preprint,3p,authoryear,lefttitle]{elsarticle}
  \journal{Icarus}
\fi

\usepackage{ulem}

\usepackage{fmtcount}
\usepackage{amsfonts}
\usepackage{siunitx}
\DeclareSIPrefix\micro{\text{\textmu}}{-3}
\usepackage{upgreek,textcomp}  
\usepackage{amssymb}
\usepackage{amsmath}
\usepackage{graphicx}
\usepackage[title]{appendix}
\usepackage{booktabs}
\usepackage[skip=4pt]{caption}
\usepackage{here}
\usepackage[utf8]{inputenc}
\usepackage[T1]{fontenc}
\usepackage{lipsum}
\usepackage{pgf}
\usepackage{textcomp}
\usepackage{tikz}
\usepackage{url}
\usepackage{xspace}
\usepackage[colorlinks=True, allcolors=RoyalBlue]{hyperref}
\usepackage[nameinlink,capitalize]{cleveref}
\crefname{section}{Sect.}{Sects.} 
\usepackage{aa_macros} 

\newcommand{\nuna}[2]{(\num{#1}) \textit{#2}}  
\newcommand{\class}[1]{#1}  

\newcommand\inputpgf[2]{{
\let\includegraphicsWithoutPath\includegraphics
\renewcommand{\includegraphics}[2][]{\includegraphicsWithoutPath[##1]{#1/##2}}
\input{#1/#2.pgf}
}}

\def\NAsteroidSpectra{48}
\def\NMeteoriteSpectra{41}
\def\NMeteorites{40}
\def\NCOMeteorites{15}
\def\NCVMeteorites{24}
\def\NCVOxAMeteorites{10}
\def\NCVOxBMeteorites{8}
\def\NCVRedMeteorites{6}
\def\NBarbarians{16}
\def\NLBarbarians{8}
\def\NAsteroidMultipleSpectra{6}
\def\NKAsteroids{11}
\def\NKAsteroidSpectra{12}
\def\NLAsteroids{20}
\def\NLAsteroidSpectra{24}
\def\NMAsteroids{8}
\def\NMAsteroidSpectra{9}
\def\NPAsteroids{2}
\def\NPAsteroidSpectra{2}
\def\NSAsteroids{1}
\def\NSAsteroidSpectra{1}

\usepackage[nolist,printonlyused]{acronym}

\begin{acronym}[MITHNEOS]
  \acro{2MASS}[2MASS]{Two Micron All Sky Survey}
  \acro{AA}[AA]{Aqueous Alteration}
  \acro{AOA}[AOA]{amoeboid olivine aggregate}
  \acro{ATLAS}[ATLAS]{Asteroid Terrestrial-impact Last Alert System}
  \acro{BIC}[BIC]{Bayesian Information Criterion}
  \acro{CC}[CC]{carbonaceous chondrite}
  \acro{CAI}[CAI]{calcium-aluminium-rich inclusion}
  \acro{CCD}[CCD]{charge-coupled device}
  \acro{CDS}[CDS]{Centre de Données astronomiques de Strasbourg}
  \acro{DAMIT}[DAMIT]{Database of Asteroid Models from Inversion Techniques}
  \acro{FoV}[FoV]{field-of-view}
  \acro{EC}[EC]{Enstatite Chondrite}
  \acro{EOC}[EOC]{Equilibrated Ordinary Chondrite}
  \acro{ECAS}[ECAS]{Eight-Color Asteroid Survey}
  \acro{EKB}[EKB]{Edgeworth-Kuiper Belt}
  \acro{EKBO}[EKBO]{Edgeworth-Kuiper Belt Object}
  \acro{ESA}[ESA]{European Space Agency}
  \acro{ESO}[ESO]{European Southern Observatory}
  \acro{IRAS}[IRAS]{Infrared Astronomical Satellite}
  \acro{FA}[FA]{Factor Analysis}
  \acro{GMM}[GMM]{Gaussian Mixture Model}
  \acro{IAU}[IAU]{International Astronomical Union}
  \acro{JAXA}[JAXA]{Japan Aerospace Exploration Agency}
  \acro{J-PAS}[J-PAS]{Javalambre-Physics of the Accelerated Universe Astrophysical Survey}
  \acro{JT}[JT]{Jovian Trojan}
  \acro{KC}[KC]{Kakangari Chondrite}
  \acro{KDE}[KDE]{Kernel Density Estimation}
  \acro{MBA}[MBA]{Main Belt Asteroid}
  \acro{LSST}[LSST]{Legacy Survey of Space and Time}
  \acro{ICL}[ICL]{Integrated Completed Likelihood}
  \acro{IDP}[IDP]{Interplanetary Dust Particle}
  \acro{IMB}[IMB]{Inner Main Belt}
  \acro{IMCCE}[IMCCE]{Institut de mécanique céleste et de calcul des éphémérides}
  \acro{JWST}[JWST]{James Webb Space Telescope}
  \acro{MC}[MC]{Mars-Crosser}
  \acro{MCFA}[MCFA]{Mixture of Common Factor Analysers}
  \acro{MCMC}[MCMC]{Markov chain Monte Carlo}
  \acro{MIR}[MIR]{mid-infrared}
  \acro{MITHNEOS}[MITHNEOS]{MIT-Hawaii Near-Earth Object Spectroscopic Survey}
  \acro{MMB}[MMB]{Middle Main Belt}
  \acro{MMR}[MMR]{mean-motion resonance}
  \acro{MPC}[MPC]{Minor Planet Center}
  \acro{NASA}[NASA]{National Aeronautics and Space Administration}
  \acro{NIR}[NIR]{near-infrared}
  \acro{NEO}[NEO]{Near-Earth Object}
  \acro{OC}[OC]{Ordinary Chondrite}
  \acro{OMB}[OMB]{Outer Main Belt}
  \acro{PC}[PC]{principal component}
  \acro{PCA}[PCA]{Principal Component Analysis}
  \acro{PDS}[PDS]{Planetary Data System}
  \acro{PPCA}[PPCA]{Probabilistic Principal Component Analysis}
  \acro{PT}[PT]{Petrologic Type}
  \acro{RC}[RC]{Rumuruti Chondrite}
  \acro{RMS}[RMS]{root mean square}
  \acro{SD}[SD]{Scattered Disk}
  \acro{SDO}[SDO]{Scattered Disk Object}
  \acro{SDSS}[SDSS]{Sloan Digital Sky Survey}
  \acro{SMASS}[SMASS]{Small Main-Belt Asteroid Spectroscopic Survey}
  \acro{SNR}[SNR]{signal-to-noise ratio}
  \acro{SsODNet}[SsODNet]{Solar system Open Database Network}
  \acro{TM}[TM]{Thermal Metamorphism}
  \acro{TNO}[TNO]{Trans-Neptunian Object}
  \acro{UOC}[UOC]{Unequilibrated Ordinary Chondrite}
  \acro{UV}[UV]{ultra-violet}
  \acro{VisNIR}[VisNIR]{visible-near-infrared}
  \acro{VISTA}[VISTA]{Visible and Infrared Survey Telescope for Astronomy}
  \acro{WISE}[WISE]{Wide-field Infrared Survey Explorer}
\end{acronym}

\begin{document}

\newcommand{\CVOxA}{CV\textsubscript{OxA}\xspace}
\newcommand{\CVOxB}{CV\textsubscript{OxB}\xspace}
\newcommand{\CVRed}{CV\textsubscript{Red}\xspace}

\ifx\destination\arxiv
\title{Spectral analogues of Barbarian asteroids among CO and CV chondrites}

\author[1,2]{Max Mahlke}

\author[3]{Jolantha Eschrig}

\author[1]{Benoit Carry}

\author[3]{Lydie Bonal}

\author[3]{Pierre Beck}
\fi

\ifx\destination\aanda
  \title{Spectral analogues of Barbarian asteroids among CO and CV chondrites}
  \subtitle{}
  \titlerunning{Asteroid Taxonomy from Cluster Analysis of Spectrometry and Albedo}
  \authorrunning{Mahlke et al.}

  \author{M.~Mahlke\inst{\ref{inst1},\ref{inst2}}\and J.~Eschrig\inst{\ref{inst3}}\and
  B.~Carry\inst{\ref{inst1}}\and L.~Bonal\inst{\ref{inst3}}\and P.~Beck\inst{\ref{inst3}}}

  \institute{Universit{\'e}
      C{\^o}te d'Azur, Observatoire de la C{\^o}te d'Azur, CNRS, Laboratoire Lagrange, France
    \label{inst1} \and
Institut d'Astrophysique Spatiale, Université Paris-Saclay, CNRS, F-91405 Orsay,
France\label{inst2}
      \and
Universit{\'e}
  Grenoble Alpes, Institut de Plan{\'e}tologie et d'Astrophysique de Grenoble, CNRS-CNES, 38000 Grenoble, France
      \label{inst3}
  }

  \date{Received date / Accepted date }
        \ifx\destination\isaanda
\abstract
{
\fi
K- and L-type asteroids are considered to be the parent bodies of CV
and CO chondrites. Spectral models of L-types invoke an enrichment in \acp{CAI}
with respect to the chondrites in the meteorite collection.
Barbarian asteroids are associated to L-type asteroids yet the
relationship between these populations is still not clear.
\ifx\destination\isaanda
}
{
\fi
 We aim to investigate the link between the K- and L-type and Barbarian
  asteroids and the CV and CO chondrites by means of spectral matching of a
  large number of reflectance spectra of objects from either population. We
  seek to identify matches based on observed rather than modelled spectral
  features.
\ifx\destination\isaanda
}
{
\fi
 We employ a matching criterion that accounts for the residuals and the
    correlation of the spectral features. The only free parameter in the
    comparison is the degree of alteration of the asteroids with respect to the
    meteorites expressed via an exponential model. We derive an
    absolute scale of similarity between the spectra using laboratory data from
  irradiation experiments.
\ifx\destination\isaanda
}
{
\fi
  \CVOxA chondrites are the best match to the asteroids, in
      particular to K-type (7 out of 11 asteroids matched) and
      Barbarians (11 out of 16). CO chondrites provide convincing matches
      for K-types (5 out of 11) and Barbarians (7 out of 16) as
      well.
    A single non-Barbarian L-type is matched to a meteorite. Only a few asteroids
  are matched to \CVOxB and \CVRed chondrites.
\ifx\destination\isaanda
}
{
\fi
 Barbarian asteroids are represented among CO and \CVOxA chondrites without
  requiring an enrichment of \acp{CAI} in the asteroids.
  Four candidate Barbarian
  asteroids are identified, three of which are classified as K-types. These
  asteroids are favourable targets for polarimetric observations. The
  discrepancy between L-type asteroids and CV and CO chondrites is likely
  related to the ambiguity of the asteroid class itself. An extension of the
  taxonomy to include polarimetric properties is required.
\ifx\destination\isaanda
}
\fi

  \maketitle

\fi

\ifx\destination\publisher
  \corref{cor}
  \cortext[cor]{Corresponding author}
  \ead{max.mahlke@oca.eu}  
\fi


\ifx\destination\arxiv
\address[1]{Universit{\'e}
  C{\^o}te d'Azur, Observatoire de la C{\^o}te d'Azur, CNRS, Laboratoire Lagrange, France}
\address[2]{Institut d'Astrophysique Spatiale, Université Paris-Saclay, CNRS, F-91405 Orsay, France}
\address[3]{Universit{\'e}
  Grenoble Alpes, Institut de Plan{\'e}tologie et d'Astrophysique de Grenoble, CNRS-CNES, 38000 Grenoble, France}

  \twocolumn[
    \begin{@twocolumnfalse}
      \maketitle
      \begin{abstract}
        
      \end{abstract}
    \end{@twocolumnfalse}
  ]
\fi
\ifx\destination\publisher
  \begin{frontmatter}
    \begin{abstract}
      
    \end{abstract}
  \end{frontmatter}
\fi

\newcommand{\sub}[1]{\ensuremath{_{\textrm{#1}}}}
\newcommand{\src}{\ensuremath{^{\textrm{\textcolor{blue}{[Citation needed]}}}}}

\section{Introduction}%
\label{sec:introduction}%
\ifx\destination\arxiv
\vspace{-0.5em}
\fi
Establishing links between meteorites and their parent asteroids is a
fundamental goal of planetary science
\citep{ForgingAnAsteGaffey1993,1995Metic..30Q.486B,LinkingAsteroiGreenw2020,ConnectingAsteDemeo2022}.
Detailed mineralogical analyses of meteorites allow us to interpret observational
features of single asteroids
\citep[e.\,g.{}][]{McCord1970Vesta,DiscoveryOfALazzar2000,MineralogicalCDeLeo2004,VisibleToNearDibb2023} and trends
among larger populations
\citep[e.\,g.{}][]{SpectroscopicSFornas2010,IdentifyingMetThomas2010,NearInfraredSDeLeo2012,CompositionalHVernaz2016,ESCHRIG2021114034,InvestigatingSEschri2022},
which in turn are used to infer their dynamical history and to constrain
models of the formation of the Solar System.

\class{K}- and \class{L}-type asteroids are rare both in terms of their
absolute number and their mass fraction with respect to the general population
of the Main Belt; combined, they represent <\SI{10}{\percent} in a given mass range
and section of the Main Belt
\citep{DeMeo2013Distribution,AsteroidTaxonoMahlke2022}. They are
observationally distinct from members of the C- and S-complex due to their
moderate albedos and colours, which typically fall between those of these latter complexes
\citep{Tedesco1989Taxonomy,DeMeo2013Distribution,2011ApJ...741...90M,Popescu2018MOVIS}.
K-type asteroids show a \SI{1}{\micro\meter} feature
associated to forsteritic olivine and, in some cases, a weak
  \SI{2}{\micro\meter} feature associated to orthopyroxene
\citep{1988Metic..23..256B,221EosARemnMothe2005,SpectroscopyOfClark2009}.
\class{L}-type asteroids exhibit a \SI{2}{\micro\meter} feature
attributed to Fe$^{2+}$-bearing spinel and a weak or fully absent \SI{1}{\micro\meter} feature attributed to Fe-rich
olivine \citep{SAsteroids387Burbin1992,AncientAsteroiSunshi2008}.
As the depths of both features vary, the spectral appearance of
both classes is continuous and members of either
class are frequently reclassified into the other one
\citep{2002Icar..158..146B,AnExtensionOfDemeo2009,AsteroidTaxonoMahlke2022}.

Based on spectral similarities, \class{K}- and L-type asteroids have been
associated primarily to two classes of \acp{CC}, namely CO and CV
\citep{1988Metic..23..256B,2001M&PS...36..245B,MineralogicalAMothe2008,SpectroscopyOfClark2009}.
These classes of anhydrous \acp{CC}
  show similar and partially overlapping distributions in
  oxygen-isotope compositions and petrographic properties,
  including similar volume percentages of chondrules, matrix, and refractory
inclusions \citep{weisberg2006systematics,2014mcp..book....1K}.
All CO and CV chondrites are of petrographic type 3.
They show the largest abundances of refractory inclusions among \acp{CC} with
\SI{13}{vol\percent} and \SI{10}{vol\percent} respectively.
More specifically, they include \acfp{CAI}
  and to a lesser degree \acp{AOA}
  \citep[$<$5vol\%,][]{AbundanceMajoEbel2016,ConstraintsOnPinto2021}. \acp{CAI} are some of the oldest components
in chondrites and consist of various minerals including melilite, forsterite and
spinel. They are believed to have condensed at high temperatures and low
pressures within the solar nebula . \acp{AOA} are micro- to millimetre sized
aggregates of forsterite, Fe-Ni metal, spinel and anorthite, among others. Most
AOAs did not undergo melting \citep{2014mcp..book...65S}.

The key differences between CO and CV chondrites are given in their whole-rock compositions,
where the latter are generally enriched in lithophile elements with respect to CO chondrites while CO are generally enriched in siderophile elements.
Furthermore, CO chondrites have considerably smaller chondrules
\citep[average diameter of \SI{0.15}{m\meter},][]{2014mcp..book....1K} compared
to CV (\SI{1}{m\meter}). Spectrally, both chondrite classes show \SI{1}{\micro\meter} features attributed to olivine and \SI{2}{\micro\meter} features attributed
to Fe$^{2+}$-bearing spinel \citep{Cloutis2012CO,Cloutis2012CV}.
The \SI{2}{\micro\meter} feature is generally absent in CO chondrites
of petrographic type $\leq$ 3.1, while the \SI{1}{\micro\meter} feature becomes more pronounced with thermal metamorphism \citep{Cloutis2012CO,ESCHRIG2021114034}.

CV chondrites are further subdivided into the
reduced \CVRed and the oxidised \CVOxA and \CVOxB
\citep{PetrographicVaMcswee1977,1997M&PSA..32R.138W}. The subgroups are based on different compositional and petrographic
properties \citep{MineralogicalAKrot1995,Cloutis2012CV}.
In particular, in comparison to \CVOxA and \CVOxB, \CVRed chondrites are characterised by (i) a
lower abundance of matrix, (ii) a higher abundance of metal, and (iii) the presence
of Ni-poor sulfides. In comparison to \CVOxB, \CVOxA are characterised by (i)
similar matrix abundance, (ii) a higher abundance of metal, (iii) the presence
of metal almost exclusively under the form of awaruite, (iv) lower Ni content of
sulfides, and (v) lower magnetic susceptibility and saturation remanence
\citep{WaterAndHeatBonal2020}. The
oxidised \class{CV} chondrites (in particular \CVOxA) generally have larger
petrographic types (>3.6) than the reduced \class{CV} and the \class{CO}
chondrites \citep{ThermalHistoryBonal2016,WaterAndHeatBonal2020}.
\citet{Cloutis2012CV} did not identify differences in the spectral appearance
between the three subgroups exceeding the variability of the spectra
within a single subgroup, while \citet{ESCHRIG2021114034} observe systematic
differences in the depths and widths of the \SI{1}{\micro\meter} and
\SI{2}{\micro\meter} features, in particular between \CVRed and \CVOxA.
\citet{ESCHRIG2021114034} further note the spectral similarity between CO
chondrites and the \CVOxA subgroup.

While it is commonly assumed that each \ac{CC} class is derived from an individual parent body,
 \citet{CvChondritesGattac2020} conclude, based on petrographic and isotopic properties, that oxidised and reduced CV chondrites have two distinct parent
bodies. Furthermore, \citet{TheRelationshiGreenw2010} suggest that
the oxidised CV subgroups and CK chondrites may have formed in the same parent
body given their similar oxygen isotopes and elemental abundances. These authors propose that the
thermally metamorphosed CK chondrites represent the core of this parent
body while
the CV chondrites form the outer shell. A possible remnant of the
partially differentiated CV-CK parent body could be the Eos family. Its members
are predominantly K-types and show a spectral variability that is consistent
with being composed of a partially differentiated ordinary-chondritic parent body
\citep{EosFamilyASDoress1998,MineralogicalAMothe2008,TheRelationshiGreenw2010}.
A further candidate family may be the
Eunomia family. Its parent body \nuna{15}{Eunomia} appears partially
differentiated with an olivine-rich composition
\citep{SpectralStudyNathue2005}.

Like K-types, L-type asteroids are commonly associated to CO and CV chondrites \citep{SAsteroids387Burbin1992,2001M&PS...36..245B}.
Using radiative transfer models, \citet{AncientAsteroiSunshi2008}
show that the spectra of \nuna{234}{Barbara}, \nuna{387}{Aquitania}, and \nuna{980}{Anacostia} can be
modelled using endmembers consisting of olivine,
 \ac{CAI}-free matrix from \CVOxA Allende, and a subtype of \ac{CAI} (fluffy type A) common in CV chondrites in addition to a slope component.
 The derived \ac{CAI} abundances are between \SI{22}{\percent} and \SI{39}{\percent}.
A large abundance of
refractory inclusions would necessitate an early formation of these asteroids,
making them the most ancient probes of the Solar System formation among the
small bodies \citep{AncientAsteroiSunshi2008}.

\citet{NewPolarimetriDevoge2018} extend this analysis to a larger sample of L-type asteroids and by including
an endmember spectrum consisting of a bulk measurement of \CVOxA Y-86751.
For the sample of 28 L-types, the required \ac{CAI} abundance
  using the same \CVOxA Allende endmember as \citet{AncientAsteroiSunshi2008} is
\SI{28(13)}{vol\percent}, while for the \CVOxA Y-86751 endmember, an abundance
of \SI{14(10)}{vol\percent} of \ac{CAI} is required to spectrally match the
sample of 28 L-type asteroids.
The highest refractory inclusion abundance observed in meteorites is
\SI{13}{vol\percent} for CO chondrites when accounting for both \acp{CAI} and
\ac{AOA}  \citep{0446bd8236e94fce91029b732260dfb7,2014mcp..book....1K}.

A different population of asteroids commonly associated to L-types are
the Barbarians. Unlike K- and L-types, this group is defined based on
polarimetric rather than spectral features. Specifically, Barbarians are defined
based on a high inversion angle ($\alpha_{\textrm{min}}>\SI{25}{deg}$)
of the negative branch of the polarimetric phase curve, as observed for
\nuna{234}{Barbara} by \citet{TheStrangePolCellin2006}.
While \citet{NewPolarimetriDevoge2018} concluded that L-types as defined by \citet{AnExtensionOfDemeo2009}
and Barbarian asteroids are identical populations, \citet{AsteroidTaxonoMahlke2022}
 show that Barbarians exhibit a spectral variability
that is larger than permissible for a single taxonomic
class.
Nevertheless,
\citet{NewPolarimetriDevoge2018} show that the Barbarian polarimetric feature
is correlated with the modelled abundance of \ac{CAI} in the asteroid,
and the authors suggest \ac{CAI} enrichment as a possible
mechanism for the large polarimetric inversion angle.
\citet{ExperimentalPhFratti2019} measure
$\alpha_{\textrm{min}}=\SI[separate-uncertainty=true]{22(1)}{deg}$
for \CVOxA Allende and
$\alpha_{\textrm{min}}=\SI[separate-uncertainty=true]{20(5)}{deg}$ for CV DaG\,521
and COs FRO\,95002 and FRO\,99040, which does not allow the reported correlation
to be confirmed or denied.
An alternative explanation is a heterogeneity of high- and low-albedo particles on the asteroid surfaces
\citep{NewCasesOfUnGilHu2008}.

In this work, we investigate the spectral match of CO and CV chondrites and
K-type,
L-type and Barbarian asteroids. One of the main focuses of our analysis
is on the question of whether a larger sample size of asteroid, and in particular
meteorite spectra, may reveal matches between the chondrites and the asteroids
without requiring an enrichment in \acp{CAI}. We further divide the CV
chondrites into the subgroups \CVOxA, \CVOxB, and \CVRed, in line with the
current interpretation of the CV class in the literature. In \cref{sec:methodology}, we
outline the sample preparation of asteroid and meteorite spectra as well as the
matching procedure. In \cref{sec:results}, we present our results.
In \cref{sec:discussion}, we draw conclusions based on the
derived similarities of the populations.

\section{Methodology}%
\label{sec:methodology}%

In this section, we first outline our methods of sample collection and preparation,
followed by a description of the matching algorithm and the derivation of an
absolute similarity scale used to identify matching pairs of asteroids and
meteorites.

\subsection{Sample selection}
\label{sec:sample}

\begin{figure*}[pt]
  \centering
  \input{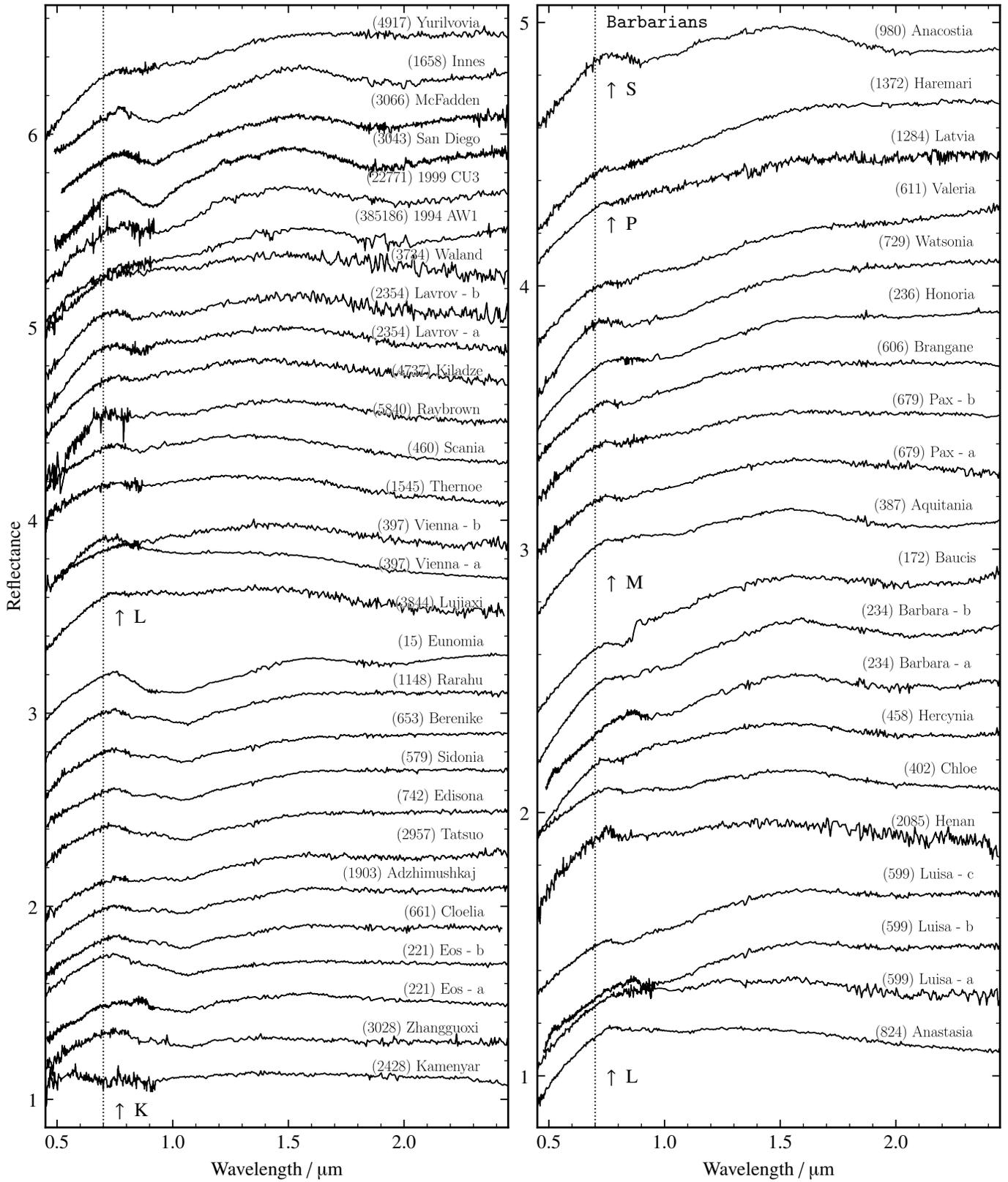}
  \caption{Reflectance spectra of non-Barbarian \class{K}- and \class{L}-type
    asteroids (left) and Barbarian
    asteroids (right). The spectra are sorted by class and decreasing
    \acs{NIR} slope. Wavelengths below \SI{0.7}{\micro\meter} (dotted
line) are excluded from the following analysis, as outlined in the text.
The spectra are shifted along the y-axis for
comparability.
Their references are given in \cref{app:asteroid_spectra}.
}
  \label{fig:klm}
\end{figure*}

\begin{figure*}[pt]
  \centering
  \input{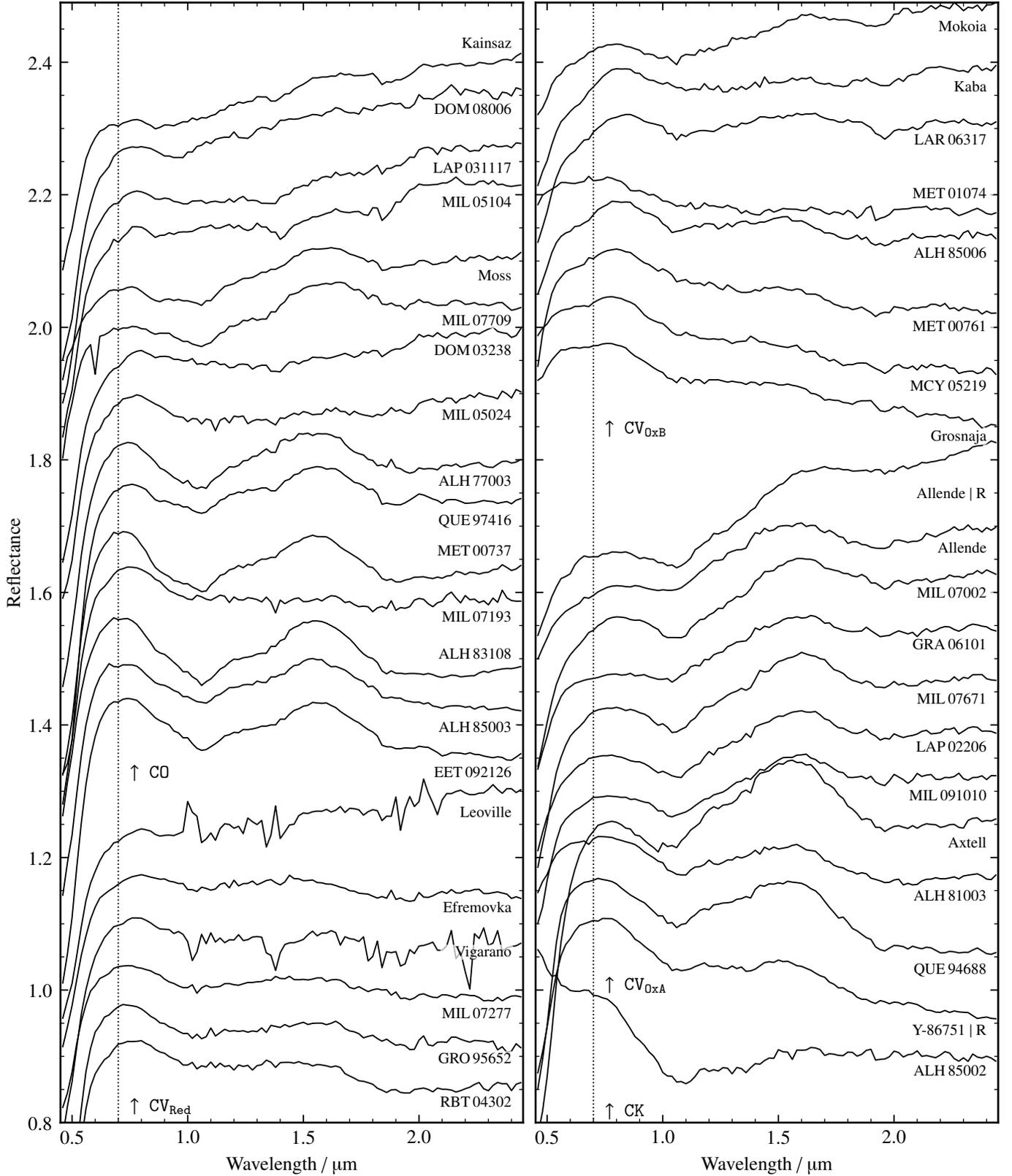}
  \caption{Reflectance spectra of \class{CK}, \class{CO}, and \class{CV}
  chondrites. The spectra are sorted by class and decreasing
  \acs{NIR} slope.
    The two spectra from the RELAB database are marked by an `R'
    beside the name of the respective meteorite.
  Wavelengths below \SI{0.7}{\micro\meter} (dotted
  line) are excluded from the following analysis.
The
spectra are shifted along the y-axis for comparability.
The spectra of the CO chondrites are from \citet{Eschrig2019CO}, and those of CV
chondrites are from \citet{Eschrig2019CV}. The measurement of the CK chondrite is unpublished (J. Eschrig).}
  \label{fig:ck_co_cv}
\end{figure*}

\subsubsection{Asteroids}
\label{sec:asteroids}

The asteroid spectra used here are compiled from various online
repositories and publications (refer to \cref{app:asteroid_spectra}). The
compilation and dataset are described in detail in
\citet{AsteroidTaxonoMahlke2022}. From this dataset, we select \ac{VisNIR}
spectra from \SI{0.45}{\micro\meter} to \SI{2.45}{\micro\meter} of asteroids
classified as \class{K}-types and \class{L}-types in
\citet{AsteroidTaxonoMahlke2022} (meaning that the probability to
be K- or L-type is larger than any other class probability) as well as of confirmed Barbarian asteroids
following the census presented in \citet{NewPolarimetriDevoge2018}.
The Barbarian asteroids are not necessarily classified as K- or
L-types in \citet{AsteroidTaxonoMahlke2022}, as is the case for S-type
\nuna{980}{Anacostia}. Spectra with low signal-to-noise ratio in
particular towards the \ac{NIR} are rejected
following visual inspection. An example is given in
\cref{fig:rejected}, where the noise towards the \ac{NIR} does not allow us to
reliably identify the presence and depth of a feature around \SI{2}{\micro\meter}. In total, there are \num{\NAsteroidSpectra}
spectra, which are shown in
\cref{fig:klm}. \Numberstringnum{\NKAsteroidSpectra} spectra belong to \num{\NKAsteroids}
\class{K}-types, \num{\NLAsteroidSpectra} spectra to \num{\NLAsteroids} \class{L}-types,
\num{\NMAsteroidSpectra} spectra belong to \num{\NMAsteroids} \class{M}-types,
\num{\NPAsteroidSpectra} spectra belong to
\num{\NPAsteroids} \class{P}-types,
and \num{\NSAsteroidSpectra} spectrum belongs to
\num{\NSAsteroids} \class{S}-type. The \num{\NBarbarians} Barbarian asteroids include
all \class{M}-, \class{P}-, and \class{S}-types, as
well as \num{\NLBarbarians} \class{L}-type asteroids.

Selecting K- and L-types based on taxonomic classifications is not
straightforward because of their spectral variability and the continuity between
the classes, making a clear separation challenging
\citep{AnExtensionOfDemeo2009,AsteroidTaxonoMahlke2022}. L-types are further
prone to misclassification as S-types and vice versa due to their \SI{2}{\micro\meter}
bands. Using a probabilistic classification scheme shows the uncertainty in the
spectral classification.
\cref{tab:meta_asteroids} gives the taxonomic classifications of the asteroids
in this study in the systems of \citet{AsteroidTaxonoMahlke2022} and Bus-DeMeo
  \citep{2002Icar..158..146B,AnExtensionOfDemeo2009}.
For spectra not classified in \citet{AnExtensionOfDemeo2009},
we used the \texttt{classy} tool\footnote{\url{https://classy.readthedocs.io}}
to classify them. In some cases, this required extrapolation of the observed spectral range.
The maximum extrapolated range represents 7.1\% of the classified spectrum and
hence we do not expect this to affect the resulting classification.

As class
definitions in \citet{AsteroidTaxonoMahlke2022} are based on Gaussian
distributions, the large spectral variability among L-types necessarily leads
to the inclusion of edge cases with significant probabilities of belonging to
other taxonomic classes, as shown in
\cref{tab:meta_asteroids}. Example objects are \nuna{1658}{Innes} and
\nuna{980}{Anacostia}.
The class probabilities could be used to
exclude asteroids from this study in case of an ambiguous classification;
however, we choose not to cut the sample based on the probabilities as the
following analysis is on a per-object basis and there is no downside to having a
potentially misclassified object in the comparison sample.

Further indicated in \cref{tab:meta_asteroids} is the Barbarian
nature of the asteroids, based on results from \citet{NewPolarimetriDevoge2018}
and \citet{TheCalernAsteBendjo2022}. Asteroids are marked with a dash if there
are insufficient polarimetric data to determine the Barbarian nature.

\Cref{fig:klm} shows the variability in spectral features and slopes of the
\class{K}- and \class{L}-types as well as of the Barbarian asteroids. The
\SI{2}{\micro\meter} band in \class{L}-types varies between prominent
(e.\,g. \nuna{234}{Barbara}) and nearly absent (\nuna{824}{Anastasia}),
while some depict a \SI{1}{\micro\meter} band (\nuna{3043}{San Diego}) and
resemble \class{S}-types. For \class{K}-types, a variability of the strength of
the \SI{1}{\micro\meter} band is well established
\citep{SpectroscopyOfClark2009}.
Among the Barbarian asteroids, we observe a
similar variability to that seen among the \class{L}-types. \nuna{824}{Anastasia} appears
blue and featureless while \nuna{980}{Anacostia} has a red slope with bands
present around \SI{1}{\micro\meter} and \SI{2}{\micro\meter}. Nevertheless, as
discussed in \citet{AsteroidTaxonoMahlke2022}, the class boundaries between
\class{K}, \class{L}, and \class{M} based on \ac{VisNIR} spectra and visual
albedos are continuous, giving rise to edge cases without a conclusive
classification.

\Numberstringnum{\NAsteroidMultipleSpectra} asteroids are represented with more
than one spectrum in the sample. We choose this to understand the systematic
uncertainty that enters into the asteroid--meteorite matching when using a single
spectrum of an individual as a reference. For example, the duplicate spectra $b$
and $c$ of \nuna{599}{Luisa} contain the same \ac{NIR} spectrum but different
visible spectra, which results in a noticeable shift of the visible feature from around
\SIrange{0.8}{1.0}{\micro\meter}.

\subsubsection{Meteorites}
\label{sec:meteorites}

A total of \num{\NMeteoriteSpectra} reflectance spectra of \num{\NMeteorites}
individual chondrites are analysed in this study,
 including
\num{\NCOMeteorites} \class{CO} and \num{\NCVMeteorites} \class{CV} spectra, the
latter of which are divided into \num{\NCVOxAMeteorites}
\CVOxA, \num{\NCVOxBMeteorites}
\CVOxB, and \num{\NCVRedMeteorites}
\CVRed, as  shown in \cref{fig:ck_co_cv}.
The majority of the spectra were presented in
\citet{Eschrig2019CV,Eschrig2019CO,ESCHRIG2021114034} and are available online in the SSHADE database
\citep{Schmitt2018SSHADE}.\footnote{\url{https://www.sshade.eu/data/experiment/EXPERIMENT_LB_20191220_001}}\textsuperscript{,}\footnote{\url{https://www.sshade.eu/data/experiment/EXPERIMENT_LB_20191220_002}}

We further add one spectrum for each of the \class{CV}\textsubscript{OxA} chondrites
Allende and Y-86751 from the RELAB database \citep[respective specimen IDs:
\texttt{MT-TJM-071} and \texttt{MP-TXH-009}]{2004LPI....35.1720P}. These spectra are
used in \citet{AncientAsteroiSunshi2008} and \citet{NewPolarimetriDevoge2018} to
study the mineralogical abundances of \class{L}-type asteroids and therefore allow us to
compare our results to those studies. We do not add
more spectra from RELAB as we aim for consistent sample treatment and
measurement conditions for the meteorite spectra in order to reduce the spectral
variability introduced by changes in observation geometry and sample properties
such as the grain size \citep{Cloutis2012CV,Cloutis2012CO,ESCHRIG2021114034}.
Finally, we acquire a spectrum of \class{CK4} chondrite ALH\,85002 to extend the
sample towards the analogue proposed by \citet{MineralogicalAMothe2008} for
\class{K}-types of the Eos family. More samples of CK chondrites
were not available to us for this work. Given the variability of
CK chondrites reported by \citet{Cloutis2012CK}, any conclusion based on a
single sample is highly tentative. Nevertheless, we choose to include this
sample based on the relationship between CV and CK chondrites
proposed by \citet{TheRelationshiGreenw2010}.

Apart from the two RELAB measurements, all reflectance spectra were acquired
with a consistent sample preparation and measurement procedure.
The chondrite samples were hand ground to a powder of
  approximately submillimetre grain size \citep{BidirectionalRGarenn2016} using
  a pestle and mortar. In contrast to the method used to obtain the RELAB
  spectra, no sieving was
  performed by \citet{ESCHRIG2021114034} in order to avoid a selection effect of
  harder-to-grind chondrite components. \citet{BidirectionalRGarenn2016}
  estimate the average grain size of hand-ground chondrite samples to
  \SIrange{100}{200}{\micro\meter}. \SI{50}{mg} of the chondrite powder was
  added to a sample holder and the surface was flattened using a spatula to
  facilitate comparison between measurements. Reflectance spectra in the
  range of \SIrange{340}{4200}{\nano\meter} were obtained at \SI{80}{\celsius}
  under vacuum to eliminate terrestrial water contamination using a
  measuring geometry of $i=\SI{0}{\degree}$, $e=\SI{30}{\degree}$.

There are two spectra available for CV
Allende, one from RELAB and one from \citet{ESCHRIG2021114034}. While the
former was measured on a \ac{CAI}-free powder of $<$\SI{38}{\micro\meter} grain size under ambient
temperatures and pressures, the spectrum in \citet{ESCHRIG2021114034} was
taken on the bulk powder using the conditions described above. These
differences in measuring conditions may partly explain the differences observed between these two
spectra. Both the removal of the \ac{CAI} component
and the decrease in grain size of the RELAB sample with respect to that
from \citet{ESCHRIG2021114034}
could explain the decrease in band depth of the
\SI{2}{\micro\meter} feature
\citep{EffectsOfHypeMustar1997,InvestigatingSEschri2022}.

\cref{fig:ck_co_cv} reveals a large degree of spectral variability between the
meteorite classes and even among samples of the same class or subclass. Both CV
and CO spectra are variable in band structure and slope. The spectra of CO
chondrites tend to have more
pronounced \SI{1}{\micro\meter} bands though there are samples such as
MIL\,07193,
which show no band at all. \citet{Cloutis2012CV} note that the petrologic parameters
used to differentiate CO and CV chondrites do not give rise to appreciable
differences in the spectra, in particular on fine-grained parent body
surfaces. Furthermore, the authors cannot establish spectral differences between CV
subtypes, while they are apparent in the sample of \citet{ESCHRIG2021114034}, as
shown in \cref{fig:ck_co_cv} and discussed in the original
publication. The
spectra of \CVOxA look similar to those of CO chondrites. Also apparent are
large differences in the visible slope, likely due to the formation of ferric
oxides as part of terrestrial
weathering \citep{MeteoriteSpectSalisb1974,1982LPSC...12.1105G,Cloutis2012CV,ESCHRIG2021114034}, and we exclude the region below
\SI{0.7}{\micro\meter} from the analysis due to this
systematic uncertainty.

\subsection{Spectral matching of asteroids and meteorites}
\label{sub:asteroidmeteoritematching}

We first define a criterion to quantify the similarity between two reflectance
spectra and then outline the assumptions we make to define an absolute scale of similarity.

\subsubsection{Similarity criterion $\Phi$}
\label{subsub:goodnessoffit_estimation}

To quantify the similarity between two reflectance spectra $\mathbf{X}$ and
$\mathbf{Y}$ consisting of $N$ datapoints $x_i$ and $y_i$ with
$i\in\{1,\dots,N\}$ sampled at the same wavelengths $\lambda_i$, we combine two
criteria presented in \citet{ModelingOfAstPopesc2012}. The first criterion
quantifies the similarity by means of the residuals $e_i$ given by $(x_i -
y_i)$,

\begin{equation}
  \label{equ:phi_res}
  \Phi_{\mathrm{res}} = \frac{1}{N}\sqrt{\sum_i^N(e_i-\bar{e})^2},
\end{equation}

where $\bar{e}$ is the mean residual value. Smaller values of
$\Phi_{\mathrm{res}}$ indicate an increasing similarity between the spectra. The
second criterion quantifies the covariance $\mathrm{cov}(\mathbf{X},
\mathbf{Y})$ of the curves,

\begin{equation}
  \label{equ:phi_cov}
  \Phi_{\mathrm{cov}} = \frac{\mathrm{cov}(\mathbf{X}, \mathbf{Y})}{\sigma_{\mathbf{X}}\sigma_{\mathbf{Y}}},
\end{equation}

where $\sigma_{\mathbf{X}}$ and $\sigma_{\mathbf{Y}}$ are the respective
standard deviations. Larger values of $\Phi_{\mathrm{cov}}$ indicate an
increasing similarity between the spectra. In particular, the correlation of the
spectra quantifies their similarity in potential absorption features. Both
criteria are combined to give the similarity criterion $\Phi$,

\begin{equation}
  \label{equ:phi}
  \Phi = \frac{\Phi_{\mathrm{cov}}}{\Phi_{\mathrm{res}}},
\end{equation}

where larger values indicate an increasing similarity between the
spectra.

Prior to the comparison, all reflectance spectra are smoothed using a
Savitzky-Golay filter \citep{1964AnaCh..36.1627S}. This filter computes a polynomial
fit to all data points in a rolling window of a user-defined width and replaces the value
at the centre of the window by the value of the fitted polynomial. By visual inspection of the smoothing,
we choose a polynomial degree of 3 and a rolling-window width of 41 data points.
For \nuna{172}{Baucis}, we use a width of 95 data points to smooth the systematic artefact around \SI{0.8}{\micro\meter}.
Finally, the spectra are resampled to a uniform
wavelength grid to ensure that the computed similarity criteria are comparable.

\subsubsection{Accounting for secondary spectral alterations}
\label{subsub:accounting}

The reflectance spectra of asteroids and meteorites are shaped
primarily by
their chemical composition and mineralogy, which are the properties we
aim to compare in asteroids and meteorites. However, in second order,
spectra are shaped by surface properties such as the regolith grain size and
porosity as well as alterations due to the
space- or the
terrestrial environment
\citep{MineralogyAndReddy2015,ReflectanceSpeClouti2018,InvestigatingSEschri2022}. These
secondary effects generally lead to a difference in the spectral appearance of
meteorites and their parent-body asteroid populations, which has to be accounted
for when establishing compositional connections
\citep{ForgingAnAsteGaffey1993,1995Metic..30Q.486B,ConnectingAsteDemeo2022}.
In particular, for carbonaceous chondrites, observing geometry and surface properties such as
grain size lead to considerable changes in the spectral appearance, that is,
larger than differences due to space weathering
\citep{Cloutis2012CO,Cloutis2012CV,IonIrradiationBrunet2014,
IonIrradiationLantz2015,IonIrradiationLantz2017,CompositionalHVernaz2016}.
However, these effects generally alter the spectral slope and
  the depth of absorption bands rather than the central wavelength
  \citep{IonIrradiationBrunet2014,Cloutis2012CV,LowPhaseSpectBeck2021}.
In this work, we therefore make the assumption that the secondary
  changes of the spectral continuum
  may be described in a joint model, for which we use the
exponential space-weathering model derived in
\citet{ModelingAsteroBrunet2006}:

\begin{figure}[t]
  \centering
  \input{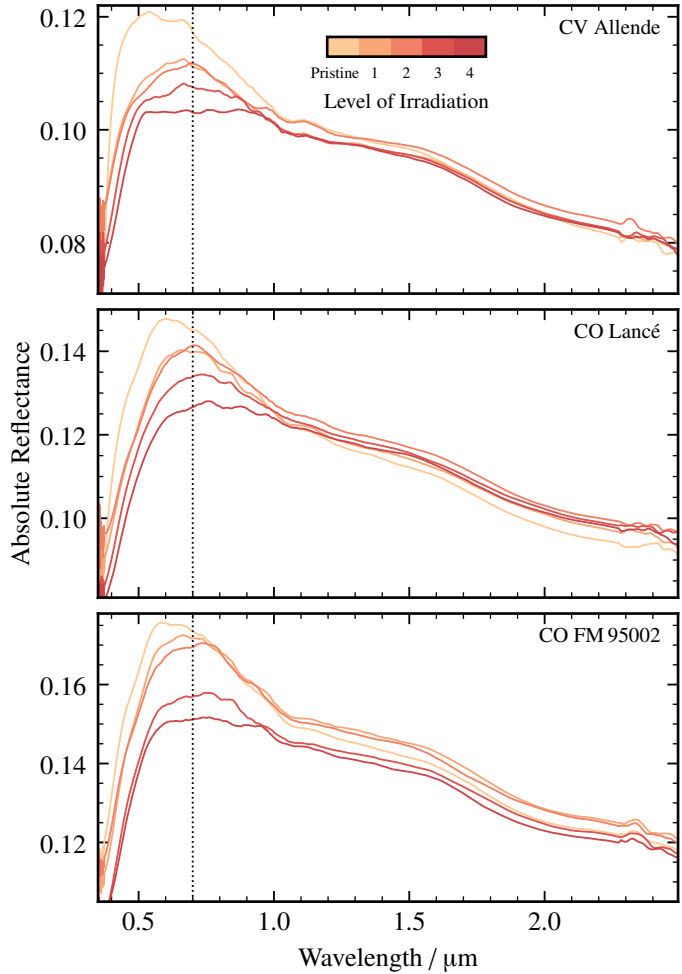}
  \caption{Reflectance spectra of pristine and irradiated carbonaceous
  chondrites from \citet{IonIrradiationLantz2017}. Wavelengths below the dotted vertical line
at \SI{0.7}{\micro\meter} are not accounted for in the analysis due to
terrestrial weathering. Data courtesy of C. Lantz.}
  \label{fig:grain_size_space_weathering}
\end{figure}

\begin{figure}[t]
  \centering
  \input{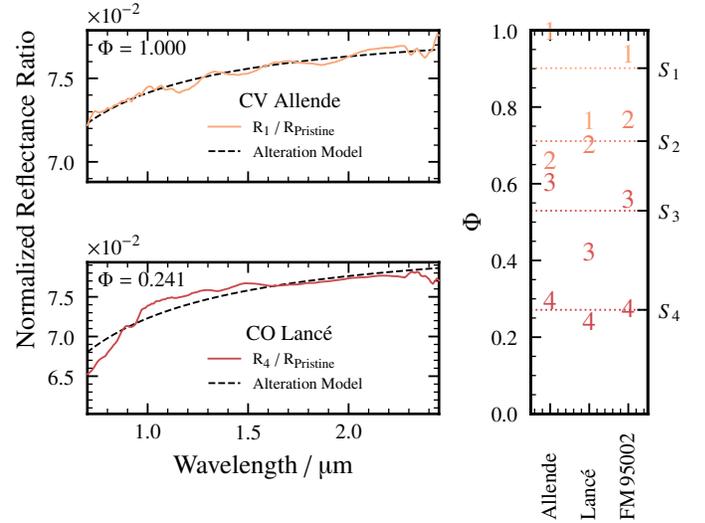}
  \caption{Evaluation of the exponential alteration model using irradiated
    meteorite samples.
Left: Ratios of irradiated to pristine spectra. Here, we show the most
(top) and the least (bottom) similar to the fitted alteration model, as
quantified by the similarity criterion $\Phi$. Right: Distribution of $\Phi$
values for all three reference meteorites and the different irradiation levels
$j$. The dotted horizontal lines indicate the mean similarity $S_j$ for each $j$.}
  \label{fig:space_weathering}
\end{figure}

\begin{equation}
  \label{equ:sw_model}
W(\lambda) = K \exp\Big(-\frac{C_{s}}{\lambda}\Big),
\end{equation}

where $W$ is the weathering function given by the ratio of the meteorite to the
asteroid spectrum, $K$ a normalising scale factor, and $C_S$ the strength
parameter of the space weathering. In the following, we refer to
this model as the alteration model in order to highlight the fact that we account for all
secondary spectral alterations with this exponential function, including but not
limited to the space weathering. The larger $C_S$, the stronger the exponential
alteration of the asteroid spectrum with respect to the meteorite
spectrum. Negative values of $C_S$ correspond to an asteroid spectrum that is
redder than the meteorite spectrum, and positive $C_S$  corresponds to a blueing of the
asteroid.
As mentioned above, we limit the spectral comparison to the range of
$\SI{0.7}{\micro\meter} \leq \lambda \leq \SI{2.45}{\micro\meter}$.

We therefore identify potential matches between
asteroids and meteorites by evaluating the similarity of their ratio to the
alteration
model given in \cref{equ:sw_model} using the similarity criterion $\Phi$
given in \cref{equ:phi}.
As we do not know the absolute values of the asteroid
reflectance spectra, we normalise the ratio to a unit $\mathbb{L}2$ norm prior
to fitting the exponential model.

\subsubsection{Quantification of similarity}
\label{subsub:estimation_of_spaceweathering}

\begin{figure*}[t]
  \centering
  \input{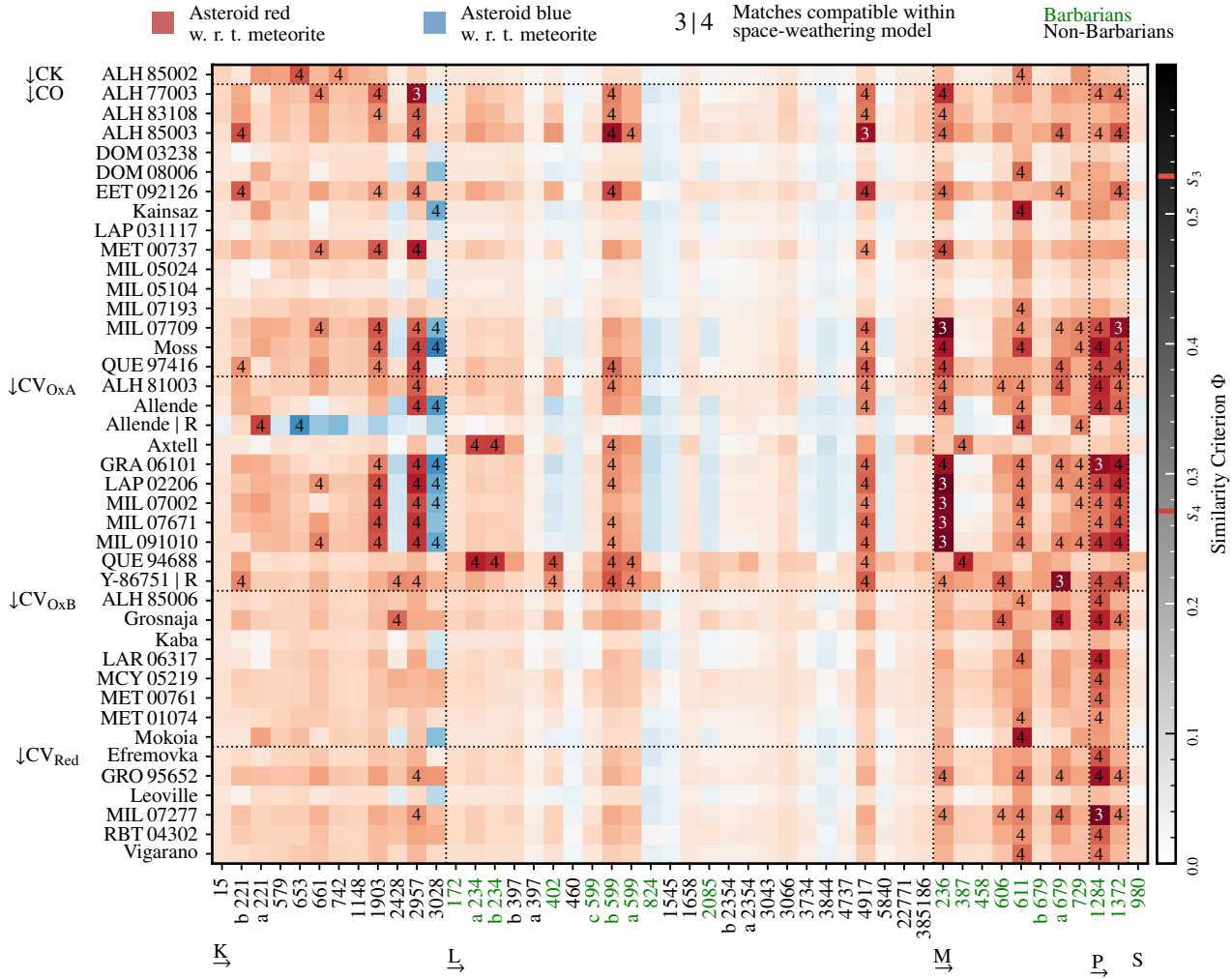}
  \vspace{-1.5em}
  \caption{Similarity $\Phi$ values
  of the alteration model to the ratio of the respective pairs
    of asteroid--meteorite reflectance spectra. Darker colours indicate
   greater similarity. Red colours show that the asteroid is reddened with
  respect to the meteorite, and blue colours indicate blueing.
  Pairs that exceed the similarity levels $S_j$ have the respective $j$
  superimposed.
  Black dotted lines
separate different classes of asteroids and meteorites. Asteroids are labelled
by their number, while Barbarian asteroids are
marked with green labels.}
\label{fig:rms}
\end{figure*}

\citet{ModelingAsteroBrunet2006} derive the model in
\cref{equ:sw_model} under the assumption that space weathering only marginally
affects absorption features. This assumption is validated for ordinary
chondritic samples and mafic silicates. We therefore have to examine whether the
model holds for CV/CO-like material as well. Furthermore,
to assess the similarity of a match, we require a scale that indicates whether
the computed values of $\Phi$ are in agreement with the assumption that
differences are induced by secondary alterations rather than by mineralogical or
compositional features. We validate the model approach and compute this scale using results from irradiation
experiments presented by \citet{IonIrradiationLantz2017} who irradiated spectra
of CV3 Allende and CO3 chondrites Lancé and FM\,95002 with He\textsuperscript{+}
ions. The \ac{VisNIR} spectra of the pristine and irradiated sample are shown in
\cref{fig:grain_size_space_weathering}. For each level of irradiation, we divide
 the irradiated spectra of each meteorite by their pristine spectrum and compute the
 $\Phi$ criterion between the obtained ratio and the fitted alteration
model. Two example fits are shown on the left-hand side of
\cref{fig:space_weathering} for meteorites Allende and Lancé, showing the
highest and the lowest similarity respectively. The right-hand side of
\cref{fig:space_weathering} shows the results for all the meteorites at the
different irradiation levels. The similarity $\Phi$ between ratio and model
decreases with increasing degree of irradiation (i.\,e. increasing degree of
space weathering) for all three meteorites. We compute the mean $\Phi$ for each
irradiation level among the three meteorites, indicated by the dotted horizontal
lines, and define them as $S_j$, where $j\in\{1, 2, 3, 4\}$. In
  \cref{fig:space_weathering} and from here on, we normalise all $\Phi$ values
  by the maximum value from this comparison ($\Phi_{\textrm{1 / Pristine}}$ of
CV Allende, top left in \cref{fig:space_weathering}) for
convenience.

The model in \cref{equ:sw_model} is used to account for additional spectral alterations.
On the other hand, the similarity scale derived in this manner only accounts for
changes due to space weathering. As such, it is a strict scale and
asteroid--meteorite pairs ruled out in this work may be considered
matches if additional discrepancies due to other spectral effects are accounted
for. Thus, the strict scale gives an increased reliability for the
matches we identify.

\section{Results}%
\label{sec:results}%

\begin{figure*}[t]
  \centering
  \input{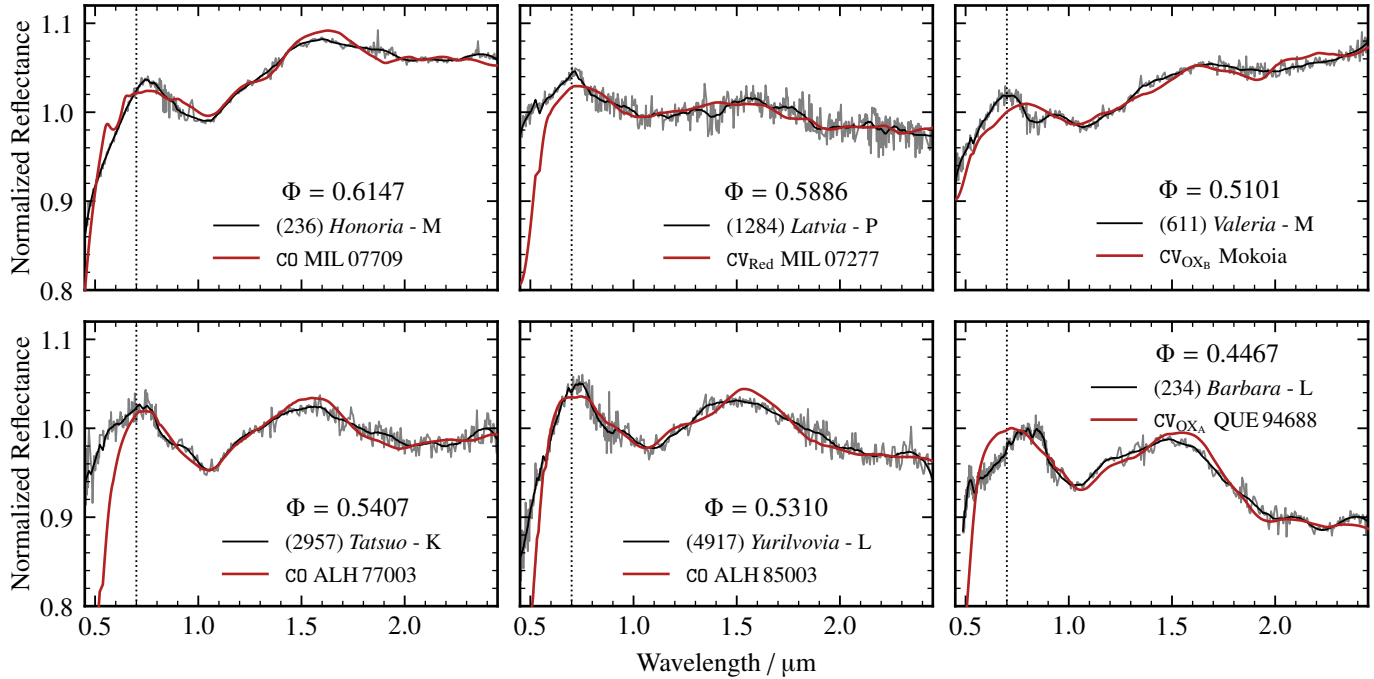}
  \vspace{-1.5em}
  \caption{Example matches of asteroids and meteorites. The asteroid spectra
    (black, solid) have been divided by the exponential
    alteration-model fitted to the ratio of the asteroid and meteorite
  spectrum. The similarity $\Phi$ between the asteroid spectra and the
  meteorite spectra (red) does not account for wavelengths below
  \SI{0.7}{\micro\meter}
  (vertical, dotted line). The light grey and thick
  black lines show the spectra
before and after smoothing, respectively. The given asteroid classes are from
\citet{AsteroidTaxonoMahlke2022}. The spectrum of
\nuna{234}{Barbara} is the one from \citet{AnExtensionOfDemeo2009}.
}
  \label{fig:best_matches}
\end{figure*}

Each of the \num{\NAsteroidSpectra} asteroid spectra in \cref{fig:klm} is
divided by each of the \num{\NMeteoriteSpectra} meteorite spectra in
\cref{fig:ck_co_cv} and the resulting ratio is fit by the alteration model in \cref{equ:sw_model}. The similarities $\Phi$ of all
model fits are shown in \cref{fig:rms}. Meteorites are aligned along the
$y$-axis, asteroids along the $x$-axis. Asteroids are indicated by their
number, which is given in green if the asteroid is a confirmed Barbarian and in
black otherwise. Darker colours in the figure indicate larger values of $\Phi$
and therefore a better description of the asteroid--meteorite ratio by the
alteration model. The cells are coloured red (blue) if the asteroid is
redder (bluer) than the meteorite. Asteroid--meteorite pairs with
$\Phi$ values above a value $S_j$ as defined in
\cref{subsub:estimation_of_spaceweathering} have the respective index $j$
superimposed in black. For these pairs, the spectral differences are well
explained by the alteration model and we consider them to be matches. No
match reaches the similarity levels $S_1$ and $S_2$. A selection of the matches
with the largest $\Phi$ is shown in \cref{fig:best_matches}, where the asteroid
spectra have been divided by the alteration model fit of the
asteroid--meteorite ratio. The spectra are normalised to minimise the
root-mean-square difference between them.

\cref{fig:rms} shows that \CVOxA chondrites have the most matches with asteroids
that
surpass the $S_4$ threshold, in particular with K-types (\num{7} out of \num{11} asteroids) and Barbarians (\num{12} out of \num{16}). For non-Barbarian L-types,
only \nuna{4917}{Yurilvovia} is a match, for \num{9} out of \num{11} chondrites.
The remaining matches for \CVOxA chondrites are almost
exclusively for confirmed Barbarians.
Only half of the L-type Barbarians are matched to chondrites, and
all the matches are to different \CVOxA chondrites than for the non-L-type
Barbarians.
\nuna{234}{Barbara}
matches Allende and QUE\,94688; the latter match is shown in the bottom-right
panel of \cref{fig:best_matches}. \nuna{402}{Chloe} and \nuna{599}{Luisa} match
QUE\,94688 as well, while non-L-type Barbarians do not show similarities to
this specific \CVOxA chondrite, with the exception of
\nuna{387}{Aquitania}. Axtell is further only matched by \nuna{387}{Aquitania}
and no other non-L-type Barbarian. However, this latter group of asteroids shows
large similarities to the remaining \CVOxA chondrites, with the
exception of \nuna{458}{Hercynia} and one spectrum of
\nuna{679}{Pax}. \nuna{980}{Anacostia} does not resemble any meteorite in this
study, which is consistent with its classification as S-type in
\citet{AsteroidTaxonoMahlke2022}.
Among K-types, \nuna{1903}{Adzhimushkaj} and
\nuna{2957}{Tatsuo} match \num{5} and \num{8} of the \num{11} \CVOxA
meteorites, respectively. \nuna{3028}{Zhangguoxi} matches \num{5} \CVOxA
chondrites while
consistently showing a bluer spectrum than all of them.

Non-L-type Barbarians further match \CVOxB and \CVRed chondrites (five out of nine),
while L- and K-types do not, with two exceptions among K-types. We note that if
an asteroid matches a \CVOxA chondrite, it tends to match a CO chondrite as
well.

CO chondrites show a dichotomy, \num{6} out of 15 have noticeably smaller mean $\Phi$
scores than the others in \cref{fig:rms}. 
Five out of \num{11} K-types and \num{7} out of 16
Barbarians are matched to CO chondrites. For non-Barbarian asteroids, only
\nuna{4917}{Yurilvovia} matches CO chondrites, most consistently with
ALH\,85003.

The CK chondrite ALH\,85002 shows the greatest similarity to K-types and
matches \nuna{653}{Berenike} and \nuna{742}{Edisona}. On the other
hand, \nuna{980}{Anacostia} is among the least similar asteroids to the ensemble
of meteorites and does not have any match, which is likely due to its pronounced
\SI{1}{\micro\meter} band.

\section{Discussion}%
\label{sec:discussion}%

Under the model assumptions outlined in \cref{sec:methodology}, we draw the following conclusions from the spectral comparison.

\subsection{Matches among CO and CV for Barbarians}

We identify several CO and CV chondrites that match Barbarian asteroids,
including \nuna{234}{Barbara}, which shows a prominent \SI{2}{\micro\meter}
band and is matched to \CVOxA chondrites Axtell and QUE\,94688. The latter match is shown in
\cref{fig:best_matches}. \nuna{387}{Aquitania} is matched to the same
meteorites, while \nuna{599}{Luisa} matches a variety of \CVOxA and CO
chondrites. These asteroids were subjects of the studies of
\citet{AncientAsteroiSunshi2008} and \citet{NewPolarimetriDevoge2018}
and were shown to match the CV endmember spectra
after adding \ac{CAI} components to the radiative transfer models.
The two meteorite spectra used in these latter works
(marked with the suffix `R' in \cref{fig:rms}) do not match \nuna{234}{Barbara}
or \nuna{387}{Aquitania} in this study either, which are the two Barbarians
with the most prominent \SI{2}{\micro\meter} bands. However, \nuna{599}{Luisa}
is considered a match to \CVOxA Y-86751.
Overall, non-L-type Barbarians (generally showing weaker
\SI{2}{\micro\meter} bands than their L-type counterparts) are more similar to
the meteorite spectra. \CVOxA chondrites are the most similar to Barbarian
asteroids, while CO chondrites further show considerable similarities to Barbarians as well.
The fact that the matched L-type Barbarians
 are associated to different \CVOxA than the non-L-type Barbarians
highlights the spectral variability in terms of the Barbarian features. This
finding
further strengthens the link to \CVOxA chondrites, as the groups display a
similar feature variability.

Barbarians \nuna{824}{Anastasia} and \nuna{980}{Anacostia} are not represented
among the CO and CV chondrites. Both asteroids represent the spectral extremes
of the Barbarian variability, both in terms of features (feature-poor versus
feature-rich) and \ac{NIR} slopes (blue versus red). A possible explanation is
therefore that the corresponding extreme endmembers of the chondrites are not present
in this study. Furthermore, \nuna{172}{Baucis} has no match.

\subsection{Possible Barbarians}
\label{sub:possible_barbarians}

Based on the spectral similarities between \CVOxA chondrites and Barbarian
asteroids, it may be worthwhile investigating the Barbarian nature of the remaining
four asteroids that show significant similarities to the chondrites, which are
the K-types
\nuna{1903}{Adzhimushkaj} , \nuna{2957}{Tatsuo},
\nuna{3028}{Zhangguoxi}, and L-type \nuna{4917}{Yurilvovia}. To our knowledge, no polarimetric phase
curves have been observed for these targets. We recommend prioritising them in
order to confirm or rule out their Barbarian nature.

\subsection{Matches with K-type asteroids inconclusive}
\label{sub:ktype_parent_bodies_inconclusive}

Apart from three asteroids (\nuna{1903}{Adzhimushkaj} , \nuna{2957}{Tatsuo}, and
\nuna{3028}{Zhangguoxi}, refer to next part), K-types only match a few
meteorites in this study, mostly CO and \CVOxA chondrites. \nuna{653}{Berenike}
and \nuna{742}{Edisona} match the only CK chondrite ALH\,85002. Both are members
of the Eos-family, for which \citet{MineralogicalAMothe2008} suggested CK
chondrites as analogue materials based on comparison of \ac{NIR} spectra. A
larger comparison of K-types and CK chondrites using a consistent meteorite
dataset could strengthen this link.

Considering the relationship between CV and CK chondrites proposed
by \citet{TheRelationshiGreenw2010}, we observe that two out of three matches
with CK ALH\,85002 are also matches to \CVOxA chondrites. However, as noted above,
our small sample size prevents us from reaching any form of conclusion here.

\subsection{Non-barbarian L-types are not parent bodies of CO and CV}
\label{sub:ltype_parent_bodies_remain_open_question}

The majority of non-Barbarian L-types
are neither matched by CO nor CV.
Even not considering possible class interlopers like
\nuna{1658}{Innes}, \nuna{3043}{San
Diego}, or \nuna{3066}{McFadden} (all three with considerable probability of
being S-types), L-types show little similarity to the chondrites in this study.
Based on the comparison here, these L-types may be ruled out as parent
bodies of CO and CV chondrites.

\subsection{Asteroid taxonomy would benefit from polarimetric measurements}
\label{sub:taxonomy_needs_to_be_extended_to_polarimetry}

The results displayed in \cref{fig:rms} clearly show that Barbarian asteroids should have their
own taxonomic class. However, their spectral variability prevents this in a
taxonomic scheme based on spectroscopy and albedo only. Polarimetric
properties should therefore be added to the asteroid
taxonomy. \citet{AsteroidTaxonoMahlke2022} highlight that polarimetric
measurements are the most promising observable to resolve compositional
ambiguities in the M-complex. The C-complex may further benefit significantly
from the addition of polarimetric measurements into the taxonomic scheme
including \ac{VisNIR} spectra and the visual albedo.

The addition of polarimetric properties to the taxonomy is currently
challenging because of the low numbers of observed asteroids. Efforts such as the
Calern Asteroid Polarization Survey \citep[CAPS,][]{TheCalernAsteBendjo2022} are
valuable steps towards a more descriptive taxonomy.

\subsection{Spectral matching for CCs requires larger sample size}
\label{sub:spectral_matching_for_cc_requires_larger_sample_size}

In \cref{sec:methodology}, we highlight the variability of reflectance spectra
of both asteroids and meteorites. The analysis in \cref{sec:results} shows that
this variability can alter the interpretation significantly: for any asteroid
and meteorite population combination (except for L-types and \CVOxB and \CVRed), we
identify both matching and non-matching pairs. Relations between
populations may therefore only be established reliably when regarding a large number
of objects.
The three spectra of \nuna{599}{Luisa} illustrate this issue: based on spectrum
(a), \nuna{599}{Luisa} does not match any meteorite in the comparison, while the
 spectra (b) and (c) are matched to 13 and 3 meteorites, respectively.

Concerning the spectra of meteorites, the variability based on the sample processing and
observation technique used is well established. Here, we note in particular that the
spectra of Allende from \citet{Eschrig2019CV} and from the RELAB database are
matched to different asteroids, in agreement with the different
sampling procedures outlined in \cref{sec:methodology}.

\section{Conclusion}
\label{sec:conclusion}

K- and L-type asteroids are rare and have been associated to various classes of
CC. For L-type asteroids, an enrichment in refractory inclusions is considered
necessary to arrive at satisfying matches with meteorite spectra. In this study,
we perform a large-scale comparison of asteroids and meteorites, focusing on K-
and L-types as well as Barbarian asteroids and on their proposed matches, the CO
and CV chondrites. The employed matching criterion $\Phi$ emphasises
the correlation between the compared reflectance spectra, which translates into
an emphasis on matching
absorption features. Spectral alterations are accounted for in a single
exponential model function. We establish matches between Barbarian asteroids and CO and
CV chondrites that do not require any additional spectral component. Four
candidate Barbarian asteroids are identified based on their matches to the
same chondrite classes as established Barbarians. For K-types and
L-types, matches among CO and CV chondrites are sparse, and we rule out the
possibility that these chondrite classes originate from non-Barbarian L-type
asteroids.


\section*{Acknowledgements}%
\label{sec:acknowledgements}%
The authors thank the referee Julia de León for the thorough and constructive review which
improved the manuscript.
The authors further thank Cateline Lantz for providing the data of the irradiation experiments.

MM acknowledges funding from the European Space Agency in the framework
of the Network Partnering Initiative. The view expressed in this
publication can in no way be taken to reflect the official opinion of the
European Space Agency.

Parts of this work have been funded by the ERC grant SOLARYS ERC-CoG2017-771691.

All (or part) of the data utilised in this publication were obtained and made
available by the MITHNEOS MIT-Hawaii Near-Earth Object Spectroscopic Survey. The
IRTF is operated by the University of Hawaii under contract 80HQTR19D0030 with
the National Aeronautics and Space Administration. The MIT component of this
work is supported by NASA grant 80NSSC18K0849.

\ifx\destination\arxiv
  \bibliographystyle{aux/arxiv}
\fi
\ifx\destination\publisher
  \bibliographystyle{aux/publisher}
  \biboptions{authoryear}
\fi

\ifx\destination\aanda
 \bibliographystyle{aux/aa} 
\fi

\bibliography{aux/bib}

\clearpage

\onecolumn

\renewcommand\thefigure{\thesection.\arabic{figure}}
\setcounter{figure}{0}
\renewcommand\thetable{\thesection.\arabic{table}}
\setcounter{table}{0}
\appendix

\section{Asteroid spectra}%
\label{app:asteroid_spectra}%
\begin{table}[h]
  \centering
  \ifx\destination\arxiv
\renewcommand{\arraystretch}{0.97}
\fi
  \captionsetup{width=1.1\textwidth}
\ifx\destination\isaanda
  \caption{References of the asteroid spectra used in this study.}
\fi
\ifx\destination\arxiv
  \caption{References of the asteroid spectra used in this study.
Each line
  refers to one spectrum. If there are several spectra of an individual
asteroid, they are denoted using (a), (b), or (c). Two references are given for
a single spectrum, they refer to the visible and \ac{NIR} parts
respectively.
Columns M22 and DM09 give the taxonomic classifications from
\citet{AsteroidTaxonoMahlke2022} and \citet{AnExtensionOfDemeo2009}
respectively.
Barbarian (Bar.) asteroids are indicated following
\citet{NewPolarimetriDevoge2018}. If the asteroid does not appear
in the polarimetric database of \citet{TheCalernAsteBendjo2022}, it is marked with a dash to indicate that the Barbarian nature is unclear.
  }
\fi
  \label{tab:meta_asteroids}
  \hspace*{-2em\relax}
  \begin{tabular}{rlllll}
	\toprule
	Number & Name & M22 & DM09 & Bar. & Reference \\
	\midrule
15 & Eunomia & K (80\%), S (20\%) & K & No & \citet{AnExtensionOfDemeo2009} \\
172 & Baucis & L (95\%) & L & Yes & \citet{NewPolarimetriDevoge2018} \\
221 & Eos & K (71\%), M (21\%), S (8\%) & Cb & No & (a) \citet{SmallMainBeltXuSh1995}, \citet{SpectroscopyOfClark2009}\\&&K&K&&(b) \citet{AnExtensionOfDemeo2009} \\
234 & Barbara & L (90\%), M (10\%) & L & Yes & (a) \citet{SmallMainBeltXuSh1995}, \citet{IrtfObservatioGietze2012}\\&&L (71\%), S (29\%)&L&&(b) \citet{AnExtensionOfDemeo2009} \\
236 & Honoria & M (87\%), P (6\%), K (4\%) & L & Yes & \citet{AnExtensionOfDemeo2009} \\
387 & Aquitania & M (85\%), L (15\%) & L & Yes & \citet{AnExtensionOfDemeo2009} \\
397 & Vienna & L (91\%), M (4\%) & L & - & (a) \citet{Bus2002TheObservations},  \citet{SpectroscopyOfClark2009}\\&&L&L&&(b) \citet{2007PDSS...51.....L,SpectroscopyOfClark2009} \\
402 & Chloe & L (88\%), M (9\%) & L & Yes & \citet{AnExtensionOfDemeo2009} \\
458 & Hercynia & M (75\%), L (13\%), S (13\%) & L & Yes & \citet{NewPolarimetriDevoge2018} \\
460 & Scania & L & L & - & \citet{AnExtensionOfDemeo2009} \\
579 & Sidonia & K (52\%), S (47\%) & K & - & \citet{AnExtensionOfDemeo2009} \\
599 & Luisa & L & L & Yes & (a) \citet{SpectralProperBinzel2001}, \citet{SpectroscopyOfClark2009}\\&&L&L&&(b) \citet{SmallMainBeltXuSh1995}, \citet{AncientAsteroiSunshi2008}\\&&M (82\%), P (9\%), S (7\%)&L&&(c) \citet{Bus2002TheObservations}, \citet{AncientAsteroiSunshi2008} \\
606 & Brangane & M (77\%), P (21\%) & L & Yes & \citet{AnExtensionOfDemeo2009} \\
611 & Valeria & M (86\%), P (14\%) & X & Yes & \citet{NewPolarimetriDevoge2018} \\
653 & Berenike & K & K & - & \citet{AnExtensionOfDemeo2009} \\
661 & Cloelia & K (88\%), M (11\%) & K & - & \citet{AnExtensionOfDemeo2009} \\
679 & Pax & M (83\%), P (6\%), S (6\%) & L & Yes & (a) \citet{AnExtensionOfDemeo2009}\\&&M (87\%), K (8\%)&L&&(b) MITHNEOS (Unpublished) \\
729 & Watsonia & M (87\%), K (8\%), P (5\%) & L & Yes & \citet{AnExtensionOfDemeo2009} \\
742 & Edisona & K (88\%), M (11\%) & K & - & \citet{AnExtensionOfDemeo2009} \\
824 & Anastasia & L & L & Yes & \citet{AnExtensionOfDemeo2009} \\
980 & Anacostia & S (62\%), L (20\%), M (19\%) & S & Yes & \citet{Bus2002TheObservations}, \citet{AncientAsteroiSunshi2008} \\
1148 & Rarahu & K (86\%), S (13\%) & K & - & \citet{AnExtensionOfDemeo2009} \\
1284 & Latvia & P (59\%), M (33\%), S (7\%) & L & Yes & \citet{Bus2002TheObservations}, \citet{2016PDSS..242.....R} \\
1372 & Haremari & P (72\%), M (27\%) & L & Yes & \citet{NewPolarimetriDevoge2018} \\
1545 & Thernoe & L & L & - & \citet{Bus2002TheObservations}, \citet{SpectroscopyOfClark2009} \\
1658 & Innes & L (60\%), S (40\%) & Sw & - & \citet{AnExtensionOfDemeo2009} \\
1903 & Adzhimush. & K (76\%), M (20\%) & K & - & \citet{AnExtensionOfDemeo2009} \\
2085 & Henan & L (91\%), M (6\%) & L & Yes & \citet{AnExtensionOfDemeo2009} \\
2354 & Lavrov & L & L & - & (a) \citet{NewPolarimetriDevoge2018}\\&&L (86\%), S (13\%)&L&&(b) \citet{AnExtensionOfDemeo2009} \\
2428 & Kamenyar & K (83\%), C (17\%) & X & - & \citet{CompositionalHVernaz2016} \\
2957 & Tatsuo & K (51\%), M (34\%), S (15\%) & K & - & \citet{AnExtensionOfDemeo2009} \\
3028 & Zhangguoxi & K & Sw & - & \citet{AnExtensionOfDemeo2009} \\
3043 & San Diego & L (85\%), S (15\%) & Sw & - & \citet{SpectralProperLucas2019} \\
3066 & McFadden & L (92\%), S (8\%) & Sw & - & \citet{2007PDSS...51.....L,TheMariaAsterFieber2011} \\
3734 & Waland & L (80\%), M (19\%) & L & - & \citet{AnExtensionOfDemeo2009} \\
3844 & Lujiaxi & L (90\%), S (6\%) & L & - & \citet{AnExtensionOfDemeo2009} \\
4737 & Kiladze & L & L & - & \citet{AnExtensionOfDemeo2009} \\
4917 & Yurilvovia & L (77\%), S (12\%), M (11\%) & L & - & \citet{NewPolarimetriDevoge2018} \\
5840 & Raybrown & L (67\%), M (27\%), S (5\%) & L & - & \citet{AnExtensionOfDemeo2009} \\
22771 & 1999 CU3 & L & S & - & \citet{AnExtensionOfDemeo2009} \\
385186 & 1994 AW1 & L & S & - & \citet{Binzel2019MITHNEOS} \\
	\bottomrule
\end{tabular}

\ifx\destination\isaanda
  \tablefoot{Each line
  refers to one spectrum. If there are several spectra of an individual
asteroid, they are denoted using (a), (b), or (c). Two references are given for
a single spectrum, they refer to the visible and \ac{NIR} parts
respectively.
Columns M22 and DM09 give the taxonomic classifications from
\citet{AsteroidTaxonoMahlke2022} and \citet{AnExtensionOfDemeo2009}
respectively.
Barbarian (Bar.) asteroids are indicated following
\citet{NewPolarimetriDevoge2018}. If the asteroid does not appear
in the polarimetric database of \citet{TheCalernAsteBendjo2022}, it is marked with a dash to indicate that the Barbarian nature is unclear.}
\fi
\end{table}

\clearpage

\section{Rejected spectrum}%
\label{app:rejected}%

\begin{figure}[h]
  \centering
  \input{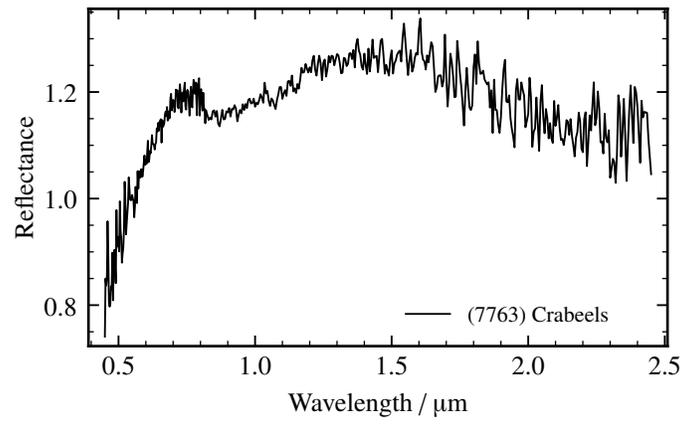}
  \caption{Example of a rejected spectrum based on the increase in the noise
  towards the \acs{NIR}.}
  \label{fig:rejected}
\end{figure}


\end{document}